\documentclass[conference]{IEEEtran}
\usepackage{graphicx}
\usepackage{algorithmic}
\usepackage{multirow}
\usepackage{subfig}
\usepackage[top=0.75in, bottom=0.75in, left=0.75in, right=0.75in]{geometry}
\ifCLASSINFOpdf
\else
\fi
\begin{document}
\title{An Approach for Spatial-temporal Traffic Modeling in Mobile Cellular Networks}
\author{\IEEEauthorblockN{Shuo Wang, Xing Zhang, Jiaxin Zhang,\\ Jian Feng and Wenbo Wang}
\IEEEauthorblockA{Wireless Signal Processing and Network Laboratory\\
Beijing University of Posts and Telecommunications,\\ Beijing, 100876, P.R. China\\
Email: wangsh@bupt.edu.cn, hszhang@bupt.edu.cn}
\and
\IEEEauthorblockN{Ke Xin}
\IEEEauthorblockA{China Telecom Corporation Limited\\ Beijing Research Institute\\
No.118 Xizhimenneidajie, Xicheng District,\\Beijing,100035,China
\\
Email: kexin@ctbri.com.cn}
}


\maketitle

\begin{abstract}
The volume and types of traffic data in mobile cellular networks have been increasing continuously. Meanwhile, traffic data change dynamically in several dimensions such as time and space. Thus, traffic modeling is essential for theoretical analysis and energy efficient design of future ultra-dense cellular networks. In this paper, the authors try to build a tractable and accurate model to describe the traffic variation pattern for a single base station in real cellular networks. Firstly a sinusoid superposition model is proposed for describing the temporal traffic variation of multiple base stations based on real data in a current cellular network. It shows that the mean traffic volume of many base stations in an area changes periodically and has three main frequency components. Then, lognormal distribution is verified for spatial modeling of real traffic data. The spatial traffic distributions at both spare time and busy time are analyzed. Moreover, the parameters of the model are presented in three typical regions: park, campus and central business district. Finally, an approach for combined spatial-temporal traffic modeling of single base station is proposed based on the temporal and spatial traffic distribution of multiple base stations. All the three models are evaluated through comparison with real data in current cellular networks. The results show that these models can accurately describe the variation pattern of real traffic data in cellular networks.
\end{abstract}


%
\IEEEpeerreviewmaketitle

\section{Introduction}
With the rapid growth of wireless terminals and the development of mobile Internet technology, the amount of traffic data in mobile cellular networks increases exponentially. Global mobile data traffic grew 81 percent in 2013 and will upsurge nearly 11-fold from 2013 and 2018 according to a forecast by Cisco \cite{cisco}. Meanwhile, because of the convergence of user behavior, the traffic distribution exhibits a strong inhomogeneous character. For example, the total traffic in dense urban area is greatly larger than that in rural areas. The traffic demand in a stadium will surge if there is a football game. In \cite{multicast}, the concept of group users' behavior (GUB) is proposed to describe the user behavior regularity in a group at a certain time and area. GUB has great impact on both the temporal and spatial traffic distributions. Therefore, the modeling of traffic patterns in different domains is of great importance for the analysis and optimization of mobile cellular networks.

Traffic volume in an area changes with time periodically which results in low traffic period and high traffic period. This is the main characteristic of traffic volume in time domain. Various models have been used for the theoretical analysis of traffic variation in cellular networks. In \cite{Oh}, the traffic profile is modeled as a periodic sinusoidal profile. It obtains the traffic profile under different loads by changing the mean and variance parameters of the sinusoidal function. This model is further modified to make it more practical in \cite{sleepwake}\cite{EEM} by adding a Poisson distributed random process to model the random fluctuations of the total traffic. In \cite{Marsan}, a trapezoidal model exploiting the family of symmetric trapezoidal curves is proposed to model the daily traffic pattern with maximum equal to 1 at the peak hour and different slopes. All these models are ideal and simplified for the tractability of theoretical analysis. There still exists a big gap between these theoretical models and real traffic data.

In spatial domain, various models also have been developed to describe the inhomogeneity of traffic distribution. Homogeneous spatial poisson point process (SPPP) and K-tier SPPP are simple and widely used distribution models for modeling of base stations' location and users' location in heterogeneous network \cite{Andrews}\cite{Dhillon}. Although SPPP model can reflect the randomness of distribution, it cannot describe the convergence of user behavior and traffic. In \cite{Ahn}, lognormal distribution model is utilized for modeling the traffic of urban area. The authors in \cite{Almeida} concluded that the traffic density follows exponential distribution by analyzing the traffic data in Lisbon and Portugal. They find that the traffic density decreases exponentially with the increase of distance from the central urban area. In \cite{Son}, the authors utilize 2-dimentianal Gaussian model for approximation of the user distribution in cellular networks. This model can generate various user distributions such as uniform distribution and hotspot clusters by adapting the parameters: the mean values and the standard deviations.

In this paper, firstly the real traffic data in a big city of China is analyzed to build traffic models of multiple base stations in temporal and spatial domain, then a combined spatial-temporal traffic modeling method is proposed for describing the traffic pattern of single base station. The parameters of proposed models in typical regions are presented. The contributions of this paper are summarized as follows:
\begin{enumerate}
\item A sinusoid superposition model is proposed for modeling the temporal traffic variation of multiple base stations in a given area based on real data from current cellular network. The main frequency components of traffic variation are presented. These components are related to user behavior.
\item Lognormal distribution is verified for spatial traffic modeling by comparing with the distribution of real traffic data. The spatial traffic distributions at both spare time and busy time are studied. The parameters of lognormal distribution in three typical regions including park, campus and central business district (CBD) are presented.
\item A spatial-temporal traffic modeling approach for generating traffic data of single base station is proposed based on the previous temporal and spatial traffic models. This model reflects both the randomness and periodicity of single base station's traffic pattern. The applications of this model may include capacity estimates, energy efficient design and other system analysis.
\end{enumerate}

The rest of the paper is organized as follows. In Section~\ref{sec:time}, a sinusoid superposition model is proposed for temporal traffic modeling of multiple base stations based on data obtained from a real cellular network. A general expression for this model is given. In Section~\ref{sec:space}, the spatial traffic distribution of real-world base stations is discussed and the parameters of the distribution model in three typical regions are presented. In Section~\ref{sec:ST}, a spatial-temporal traffic modeling approach is proposed for describing the traffic pattern of single base station more practically. Finally, conclusions are drawn and future works are discussed in Section~\ref{sec:conc}.
\section{Temporal traffic modeling of multiple base stations in real network}
\label{sec:time}
\subsection{Sinusoid superposition modeling method}
The main characteristic of traffic in cellular networks is the high and low tide effect caused by the typical day-night behavior of users. During daily hours, users commute from residential areas to office buildings with their mobile terminals, while moving towards the opposite direction at night. This pattern of user behavior results in high daytime traffic volume and low nighttime traffic volume in office districts, and the opposite regularity in residential areas.

\begin{figure}[!t]
\centering
\includegraphics[width= 8cm]{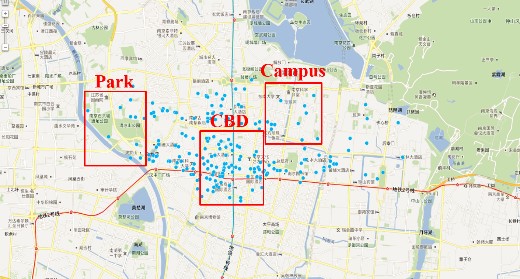}
\caption{Map of BS locations in a current cellular network}
\label{fig_1}
\end{figure}

In order to discover the regularity of traffic variation in mobile cellular networks, the real traffic data from a mobile operator in one big city of China is analyzed. The data contains information of 185 base stations from one base station controller (BSC) during three weeks ranging from 2012/9/3 to 2012/9/23 in a $6km \times 2.5km$ area as shown in Fig.~\ref{fig_1}. The recording period of the collected data is 5 minutes. At one time, the number of recorded sectors ranges from 79 to 184. The recorded information is the total traffic volume of each base station. The unit of traffic volume is Byte. The research area is located in dense urban area and contains three typical regions: park, university campus and CBD. The real data sheet of current cellular network is shown in Table~\ref{data}.

\begin{table}[!t]
\renewcommand{\arraystretch}{1.3}
\caption{Real data sheet in current network}
\label{data}
\centering
\begin{tabular}{|c|c|c|c|c|}
\hline
\multirow{2}{*} {Time} & {BS }  &\multirow{2}{*}{Longitude} & \multirow{2}{*}{Latitude}& {Traffic} \\
 & Number & & & Volume\\\hline
{2012/9/3  } & \multirow{2}{*}{BS\_1} & \multirow{2}{*}{118.7511111} & \multirow{2}{*}{32.05305556} & \multirow{2}{*}{25499860} \\
0:00 & & & & \\\hline

{2012/9/3  } & \multirow{2}{*}{BS\_2} & \multirow{2}{*}{118.7558333} &\multirow{2}{*}{32.04833333} & \multirow{2}{*}{42759785} \\
0:00 & & & & \\\hline

{...} & ... & ... & ... \\\hline
{2012/9/3} & \multirow{2}{*}{BS\_1} & \multirow{2}{*}{118.7511111} & \multirow{2}{*}{32.05305556} & \multirow{2}{*}{26912606} \\

0:05 & & & & \\\hline
{2012/9/3} & \multirow{2}{*}{BS\_2} & \multirow{2}{*}{118.7558333} & \multirow{2}{*}{32.04833333} & \multirow{2}{*}{38694353} \\

0:05 & & & & \\\hline
{...} & ... & ... & ... & ... \\\hline
\end{tabular}
\end{table}

For the purpose of mining the characteristics of overall traffic pattern, we analyze the total traffic volume of all the base stations in the research area in time domain and frequency domain, respectively. The steps of our modeling method are as follows:

Step 1: Based on the geographical information of the collected data, the whole research area is divided into typical regions. The data of base stations in each region are selected. The data contain information of time, base station coordinates and traffic volume. The geographical coordinates of base stations are converted to plane coordinates.

Step 2: For a selected region, the variation regularity of average traffic volume is analyzed. Then the main frequency components of its periodic variation are obtained via fast Fourier transform (FFT).

Step 3: Under the metric of coefficient of determination, we use as few number of frequency components as possible to guarantee both the accuracy and universality of the model. The more frequency components the model contains, the more accurate the model is for specific traffic data, but the complexity also increases and it cannot be applied to other traffic data.

Step 4: The temporal traffic model is obtained by data fitting. The coefficient parameters in our model are determined by using the MATLAB curve fitting tool.

Based on these steps, we propose a sinusoid superposition model to fit the real traffic data. The model is given by\\
\begin{equation}
V(t) = {a_0} + \sum\limits_{k = 1}^n {{a_k}\sin ({\omega _k}t + {\varphi _k})}
\end{equation}
Where $ V(t) $  is the total traffic volume of all base stations in research area, $ {a_0} $  is the average traffic volume during a period of time, $ {\omega _k} $  is the frequency components of traffic variation, $ {a_k} $  and $ {\varphi _k} $  are the amplitudes and phases respectively, $ n $ is the number of frequency components.

\begin{figure}[!t]
\centering
\includegraphics[width= 6cm]{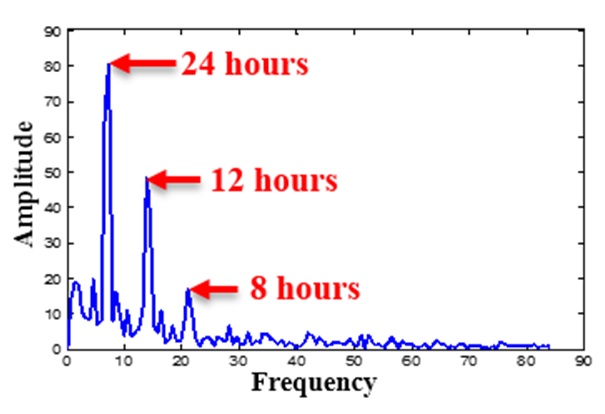}
\caption{Frequency components of real traffic data}
\label{fig_2}
\end{figure}

Applying the given method to the real data, the FFT result of real data is shown in Fig.~\ref{fig_2}. It clearly illustrates that the total traffic volume of all base stations in the given area changes mainly in three frequency components: ${\pi  \mathord{\left/
 {\vphantom {\pi  {12}}} \right.
 \kern-\nulldelimiterspace} {12}}$, ${\pi  \mathord{\left/
 {\vphantom {\pi  {6}}} \right.
 \kern-\nulldelimiterspace} {6}}$ and ${\pi  \mathord{\left/
 {\vphantom {\pi  {4}}} \right.
 \kern-\nulldelimiterspace} {4}}$, which correspond to the period of 24 hours, 12 hours and 8 hours respectively. This phenomenon is related to user social behavior, because mobile users tend to have repetitive behaviors in periods of one day, half a day and working hours. By containing these main frequency components in our model, the main characteristics of traffic variation in time domain can be represented. Therefore, a good balance between the accuracy and the universality of the model is achieved.
\subsection{Performance evaluation}
For the purpose of assessing the temporal model, we use coefficient of determination, denoted ${R^2}$, as the metric of the accuracy of the proposed model. It indicates how well data fit a statistical model. The calculation of it is expressed as follows:\\
\begin{equation}
{R^2} = \frac{{SSR}}{{SSE}} = \frac{{\sum\nolimits_{i = 1}^n {{{({{\hat y}_i} - \bar y)}^2}} }}{{\sum\nolimits_{i = 1}^n {{{({y_i} - \bar y)}^2}} }}
\end{equation}
Where ${y_i}$ is the  ${i}$th element in original data set, ${{\hat y}_i}$  is the  ${i}$th element in fitting data set, ${\bar y}$  is the mean value of original data obtained by $\bar y = \frac{1}{n}\sum\limits_{i = 1}^n {{y_i}} $ and ${n}$  is the number of elements in original data set. ${SSR}$  is the regression sum of squares and  ${SSE}$  is the total sum of squares. The closer to 1 the coefficient of determination is, the better the fitting result is.

Since the traffic changing pattern in the whole research area has three main frequency components, we use a three-order sinusoid waveform to fit the real traffic data. The fitting result can be expressed as:
\begin{equation}
\begin{array}{l}
V(t) = 173.29 + 89.83 * \sin \left( {\frac{\pi }{{12}}t + 3.08} \right) + \\52.6 * \sin \left( {\frac{\pi }{6}t + 2.08} \right)
 + 16.68 * \sin \left( {\frac{\pi }{4}t + 1.13} \right)
\end{array}
\end{equation}

\begin{figure}[!t]
\centering
\includegraphics[width= 6cm]{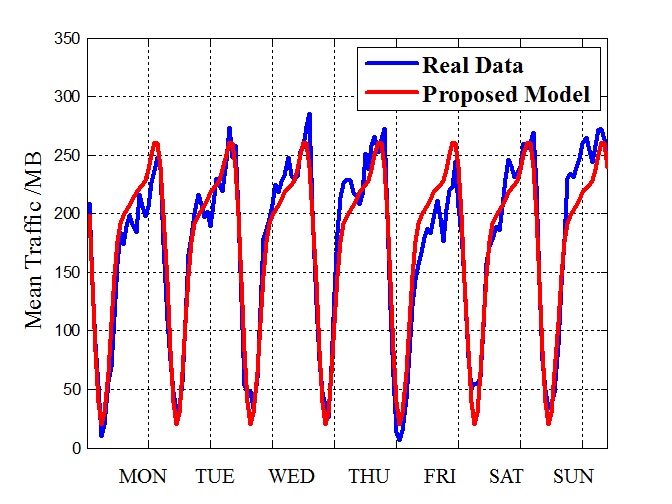}
\caption{Comparison between proposed model and real data in whole area}
\label{fig_3}
\end{figure}

Fig.~\ref{fig_3} shows the comparison between the proposed temporal traffic model and real data in a current network. The ${R^2}$ value of the proposed model is 0.9468, which means the model can fit the real data quite well. From this figure, we can also see that the variation of traffic exhibits strong periodicity and the difference of traffic volume between weekdays and weekends is not obvious.

To find out the appropriate number of frequency components of the proposed model for typical regions, we will examine the traffic data in the three following typical regions respectively: park, campus and CBD regions. The characteristics of traffic pattern in these regions are not exactly the same because the human behavior in these areas are different.
\subsubsection{Park}

\begin{figure}[!t]
\centering
\includegraphics[width= 9.2cm]{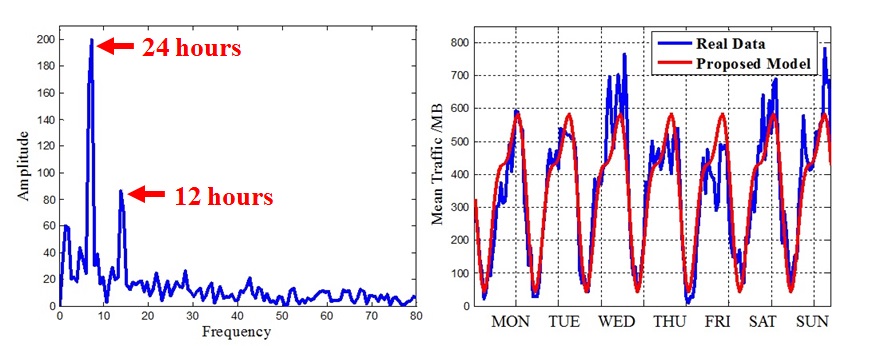}
\caption{(Left) Frequency Components of traffic in park region. (Right) Comparison between proposed model and real data in park region}
\label{park_sin}
\end{figure}

The traffic data of park region in frequency domain is shown in Fig.~\ref{park_sin} (Left). It clearly shows that the mean traffic volume of the base stations in park region changes mainly in two frequency components: ${\pi  \mathord{\left/
 {\vphantom {\pi  {12}}} \right.
 \kern-\nulldelimiterspace} {12}}$ and ${\pi  \mathord{\left/
 {\vphantom {\pi  {6}}} \right.
 \kern-\nulldelimiterspace} {6}}$, which means the traffic pattern has periods of 24 hours and 12 hours respectively. Utilizing this character, we use a two-order sinusoid waveform to fit the real traffic data. The fitting result can be expressed as:
\begin{equation}
\begin{array}{l}
{V_{park}}(t) = 351.06 + 222.7*sin\left( {\frac{\pi }{{12}}t + 3.11} \right) + \\
96.24*sin\left( {\frac{\pi }{6}t + 2.36} \right)
\end{array}
\end{equation}

Fig.~\ref{park_sin} (Right) shows the comparison of mean traffic volume between the proposed model and real data in park region. The ${R^2}$ value of the proposed model is 0.8209, which means the given model can fit the real data with great accuracy. From this figure, we can also see that the mean traffic volume at weekends is higher than that at weekdays in park region. The reason is there are more people going to park to enjoy their weekends than during weekdays.

\subsubsection{Campus}
\begin{figure}[!t]
\centering
\includegraphics[width= 9.2cm]{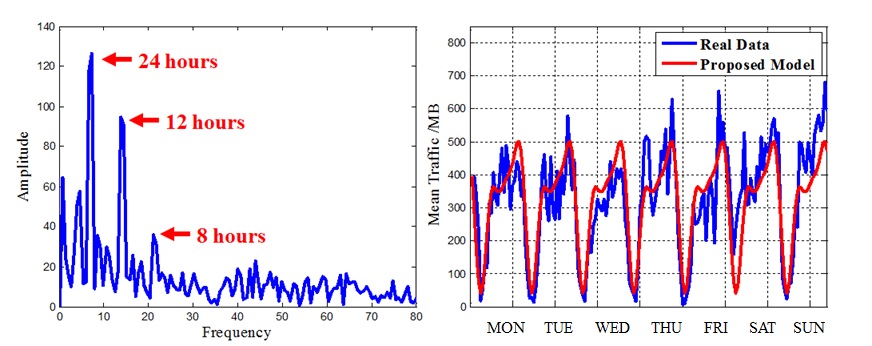}
\caption{(Left) Frequency Components of traffic in campus region. (Right)Comparison between proposed model and real data in campus region}
\label{campus_sin}
\end{figure}

The traffic data of campus region in frequency domain is illustrated in Fig.~\ref{campus_sin} (Left). In this figure, we can see the three main frequency components of the mean traffic volume of the base stations in campus region are:
${\pi  \mathord{\left/
 {\vphantom {\pi  {12}}} \right.
 \kern-\nulldelimiterspace} {12}}$, ${\pi  \mathord{\left/
 {\vphantom {\pi  {6}}} \right.
 \kern-\nulldelimiterspace} {6}}$ and ${\pi  \mathord{\left/
 {\vphantom {\pi  {4}}} \right.
 \kern-\nulldelimiterspace} {4}}$. The corresponding periods are 24 hours, 12 hours and 8 hours respectively. Utilizing this character, we use a three-order sinusoid waveform to fit the real traffic data. The fitting result can be expressed as:
\begin{equation}
\begin{array}{l}
{V_{campus}}(t) = 323.04 + 148.3*sin\left( {\frac{\pi }{{12}}t + 2.98} \right) + \\
109.4*sin\left( {\frac{\pi }{6}t + 2.15} \right) + 38.43*sin\left( {\frac{\pi }{4}t + 1} \right)
\end{array}
\end{equation}

The comparison of mean traffic volume between the sinusoid superposition model and real data in campus region is shown in Fig.~\ref{campus_sin} (Right). The accuracy of the proposed model, ${R^2}$ value, is 0.7. This figure also illustrates that the traffic in campus region has higher variability than that in park region.

\subsubsection{CBD}
\begin{figure}[!t]
\centering
\includegraphics[width= 9.2cm]{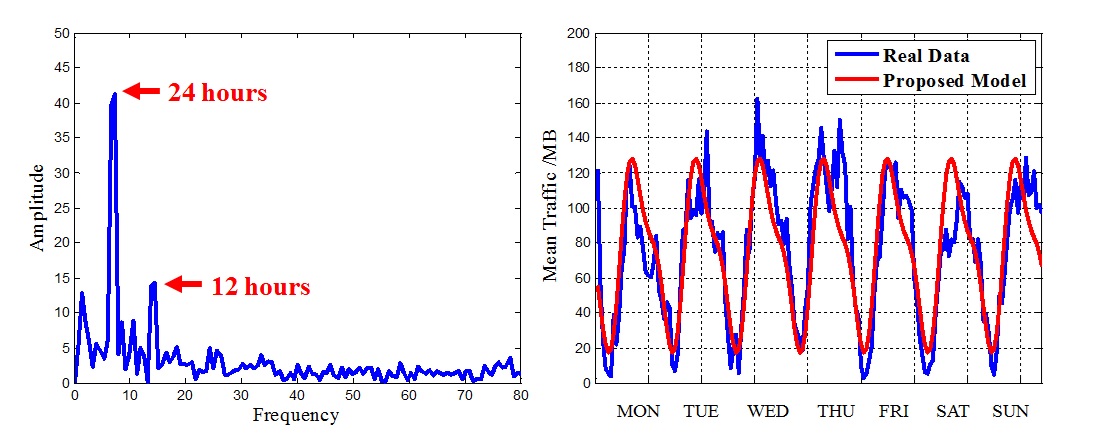}
\caption{(Left) Frequency Components of traffic in CBD region. (Right)Comparison between proposed model and real data in CBD region}
\label{cbd_sin}
\end{figure}

The traffic data of CBD region in frequency domain is shown in Fig.~\ref{cbd_sin} (Left). From this figure, we can see that the mean traffic volume of the base stations in CBD region has two main frequency components:
${\pi  \mathord{\left/
 {\vphantom {\pi  {12}}} \right.
 \kern-\nulldelimiterspace} {12}}$ and ${\pi  \mathord{\left/
 {\vphantom {\pi  {6}}} \right.
 \kern-\nulldelimiterspace} {6}}$. The corresponding periods are 24 hours and 12 hours respectively. Applying this property in our model, we use a two-order sinusoid waveform to fit the real traffic data. The fitting result can be expressed as:
\begin{equation}
\begin{array}{l}
{V_{cbd}}(t) = 75.72 + 47.52*sin\left( {\frac{\pi }{{12}}t - 2.56} \right) + \\
16.71*sin\left( {\frac{\pi }{6}t + 1.45} \right)
\end{array}
\end{equation}

The comparison of mean traffic volume between the sinusoid superposition model and real data in CBD region is shown in Fig.~\ref{cbd_sin} (Right). The accuracy of the proposed model, ${R^2}$ value, is 0.7374. This figure also illustrates that the traffic volume at weekdays is slightly higher than at weekends in CBD region.

From the above discussion, it is shown that the sinusoid superposition model can be applied universally with relatively high accuracy. The traffic volume in real networks changes at periods of 24 hours, 12 hours and 8 hours, but the number of periods varies in different type of regions.
\section{Spatial traffic modeling of multiple base stations in real network}
\label{sec:space}

With the evolution of network architecture, the traffic distribution patterns change as well. The existing traffic models have their deficiencies. For instance, SPPP and Gaussian distribution models are ideal model used for theoretical analysis, but they can hardly match the traffic data in real cellular networks. To accomplish network simulation and design of new communication system, a model which can reflect the traffic pattern of real current network is needed. Thus, we try to find the spatial characteristic of traffic from real traffic data in a current cellular network. The authors in\cite{U}\cite{Lee} conclude that the spatial inhomogeneity of traffic in cellular networks can be described by a lognormal distribution. The base stations and traffic are not uniformly distributed in real network because of the convergence of user social behavior. As a result, many hotspots where the user density is relatively high exist in current cellular network. In different type of areas, such as dense urban area and rural area, the spatial distributions of base stations are different because of the existence of hotspots in dense urban area\cite{BS}. Therefore, we want to investigate whether the traffic distribution of different type of areas is also impacted by hotspots. Whether we can model the traffic distribution of different type of areas with a unified model? In this section, firstly we analyze the spatial traffic distribution before and after removing hotspots respectively to find out what influence of the hotspots is on the spatial traffic distribution model. Then the spatial traffic distributions in three typical regions are modeled including park, campus and CBD. The parameters of the traffic model in each typical region are presented and compared. The traffic distribution at spare time and at busy time are analyzed respectively. The traffic data we analyze in this section is the same as in section~\ref{sec:time}.

\subsection{Spatial modeling of traffic considering hotspots}
In current cellular networks, operators improve network capacity by deploying more small base stations in the area where the traffic volume is high. This results in the inhomogeneity of mobile networks. In this paper, the base stations whose distances between each other are less than 150m are considered as a hotspot, or a cluster. Firstly, we select the traffic data of all base stations in the research area at a certain hour during 21 days. For example, the traffic volume of all base stations at 7 o'clock each day are selected to construct a data set. This data set contains the traffic volume of each base station at 7 o'clock in 21 continuous days. The traffic volume at each hour is the sum of traffic during that hour. Secondly, the probability density function (pdf) of the traffic volume is computed and lognormal distribution is used to fit the real data. The lognormal distribution parameters at the given time are then acquired and used to generate modeling traffic data. Then the probability density functions of real traffic data and modeling data are compared using bar graph. The probability density function of lognormal distribution is:
\begin{equation}
{f_X}(x;\mu ,\sigma ) = \frac{1}{{x\sigma \sqrt {2\pi } }}{e^{ - \frac{{{{(lnx - \mu )}^2}}}{{2{\sigma ^2}}}}},(x > 0)
\end{equation}
\subsubsection{Before removing hotspots}
\begin{figure}[!t]
\centering
\includegraphics[width= 9cm]{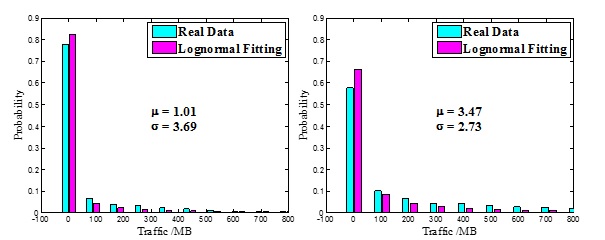}
\caption{(Left) Spatial distribution of traffic before removing hotspots at 2-4 o'clock. (Right) Spatial distribution of traffic before removing hotspots at 17-19 o'clock}
\label{fig_4}
\end{figure}

Using the method described above, we analyze the traffic data of all base stations without removing hotspots at 2 to 4 o'clock in the morning and at 17-19 o'clock in the evening. The different hours represent the traffic during spare time and busy time respectively. The probability distribution of real data and lognormal distribution fitting data at different times are shown in Fig.~\ref{fig_4}. It illustrates that lognormal distribution can be used to modeling the distribution of real traffic data quite well. The mean value $ \mu $  of the distribution in spare time is lower than in busy time.
\subsubsection{After removing hotspotss}
\begin{figure}[!t]
\centering
\includegraphics[width= 9cm]{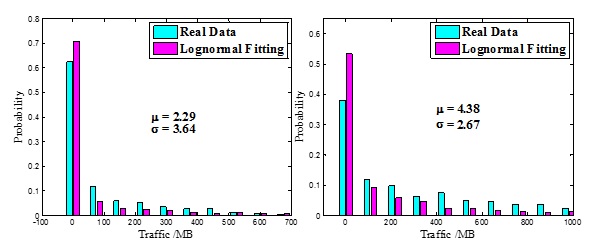}
\caption{(Left) Spatial distribution of traffic after removing hotspots at 2-4 o'clock. (Right) Spatial distribution of traffic after removing hotspots at 17-19 o'clock}
\label{fig_5}
\end{figure}

The spatial model of base stations in current networks is related to clusters \cite{BS}, or hotspots. That is to say, the spatial models of base stations before and after removing clusters are different. Whether the spatial distribution of traffic is also affected by hotspots? In order to answer this question, we find out the hotspots in the real data and remove the traffic data of them to analyze the distribution of traffic without hotspots. The probability distribution of real data without hotspots and lognormal distribution fitting data at spare time and busy time are shown in Fig.~\ref{fig_5}. It illustrates that lognormal distribution can still be used to modeling the distribution of real traffic data without hotspots quite well. The mean value of the distribution in spare time is lower than in busy time. Compared to the traffic distribution before removing hotspots, the mean value $\mu$ of the distribution after removing hotspots is greater. This is because the density of base stations is higher in hotspots, so the average traffic of each base station is lower. In other words, the average traffic of macro base stations is higher than that of small base stations.
\subsection{Spatial modeling of traffic in Typical regions}
Because of user social behaviors, user and traffic distributions are significantly related to region types. Users tend to have similar behaviors in office buildings, residential areas and so on. From s current cellar network in a city as shown in Fig.~\ref{fig_1}, we select three typical regions: park, campus and central business district, and analyze their distribution parameters respectively.
\subsubsection{Park}
\begin{figure}[!t]
\centering
\includegraphics[width= 9cm]{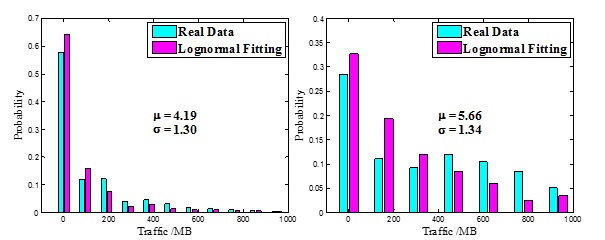}
\caption{(Left) Spatial distribution of traffic in park region at spare time. (Right) Spatial distribution of traffic in park region at busy time}
\label{fig_6}
\end{figure}

In park region, the spatial distribution of traffic also approximates to lognormal distribution at different time as shown in Fig.~\ref{fig_6}. After comparing the parameters of the traffic distribution in park region at different time, we find that the mean value $\mu$  changes with time periodically as the analysis in section~\ref{sec:time} while the standard deviation $\sigma$ is around 1.3 both at spare time and busy time.
\subsubsection{Campus}
\begin{figure}[!t]
\centering
\includegraphics[width= 9cm]{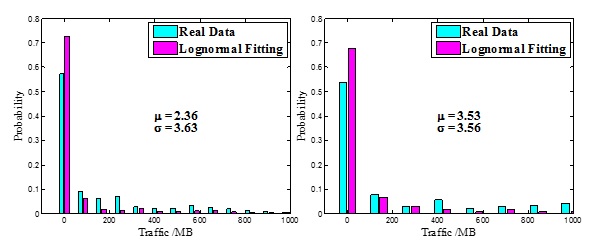}
\caption{(Left) Spatial distribution of traffic in campus region at spare time. (Right) Spatial distribution of traffic in campus region at busy time}
\label{fig_7}
\end{figure}

In campus region, lognormal distribution can still be used to model the traffic distribution as shown in Fig.~\ref{fig_7}. The mean value $\mu$ in campus region also changes with time periodically. Moreover, the mean value in campus region is lower than that in park region at the same time in a day. This is because the density of base stations in campus region is higher than the density in park region, so the average traffic each base station handles in campus region is lower than that in park region. The standard deviation $\sigma$ in campus region only correlated with region type and is around 3.6.
\subsubsection{Central business district}
\begin{figure}[!t]
\centering
\includegraphics[width= 9cm]{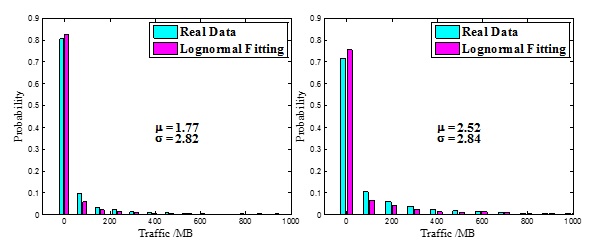}
\caption{(Left) Spatial distribution of traffic in CBD region at spare time. (Right) Spatial distribution of traffic in CBD region at busy time}
\label{fig_8}
\end{figure}

Fig.~\ref{fig_8} indicates that lognormal distribution can also match the traffic distribution of real data in CBD region. Likewise, the mean value $\mu$ in CBD region also changes with time periodically. Besides, the mean value in campus region is the lowest in the three typical regions discussed above at the same time in a day. This is because the density of base stations in CBD region is the highest, so the average traffic each base station handles in this region is lower than that in both park and campus regions. The standard deviaton $\sigma$ in CBD region is around 2.8 both at spare time and busy time.

In summary, lognormal distribution can be used to model the spatial distribution of traffic in real networks. The spatial distribution of traffic is determined by user's behavioral patterns. Lognormal distribution reflects the differentiation of traffic demand. Base stations in current cellular network provide both low data rate services such as voices and high data rate services such as high definition videos, and most part of the services are low data rate services. This fact is in accordance with the long-tailed characteristic of lognormal distribution. Finally, the empirical values of $\sigma$ in the distribution model is obtained in typical regions including park, campus and CBD which are concluded in Table~\ref{values}.
\begin{table}[!t]
\renewcommand{\arraystretch}{1.3}
\caption{Empirical values of $\sigma$ in typical regions}
\label{values}
\centering
\begin{tabular}{|c|c|c|c|}
\hline
{Regions} &{Park} & {Campus}&{CBD} \\\hline
{$\sigma$} & 1.3 & 3.6 & 2.8 \\\hline
\end{tabular}
\end{table}
\section{Traffic modeling of single base station in typical area}
\label{sec:ST}
\begin{figure}[!t]
\centering
\includegraphics[width= 9cm]{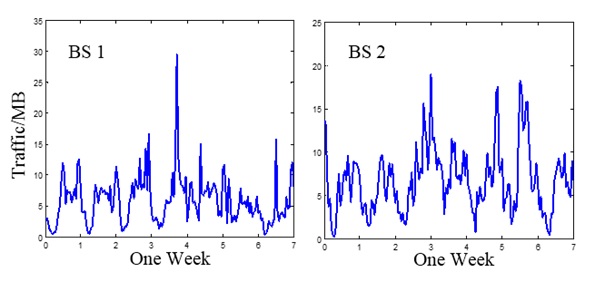}
\caption{Traffic variations of two base stations in one week}
\label{fig_9}
\end{figure}

In system analysis, it is necessary to build a traffic model for single base station. Fig.~\ref{fig_9} shows the traffic variations of two real-world base stations in one week. From this figure, we can see that the traffic variation of one single base station is quite random and has no obvious regularity. The traditional models such as sinusoid traffic model cannot reflect the randomness of single base station's traffic. In this section, we will present a spatial-temporal traffic modeling approach to model the traffic pattern of single base station more accurately.
\subsection{Spatial-temporal traffic modeling approach}
In the previous sections, the temporal and spatial traffic models of multiple base stations are presented. Based on these models, a new spatial-temporal traffic modeling approach is proposed. It considers both the periodicity and randomness of single base station's traffic pattern.

As for a typical region contains $N$ base stations, the proposed modeling method comprises three steps:

Step 1: Calculate the average traffic volume of all base stations in the region using the sinusoid superposition model proposed in section~\ref{sec:time}. The process can be expressed as:
\begin{equation}
m(t) = \frac{1}{N}\left( {{a_0} + \sum\limits_{k = 1}^3 {\sin (\frac{{k\pi }}{{12}}t + {\varphi _k})} } \right)
\end{equation}

Step 2: According to the empirical value of $\sigma$ given in section~\ref{sec:space}, compute the parameter $\mu$ of lognormal distribution at each time with the following expression:
\begin{equation}
\mu (t) = \log [m(t)] - \frac{1}{2}{\sigma ^2}
\end{equation}

This expression is obtained based on the relationship between the mean and variance of a lognormal random variable and the mean and standard deviation of the associated normal distribution expressed as:
\begin{equation}
\mu  = \log ({{{m^2}} \mathord{\left/
 {\vphantom {{{m^2}} {\sqrt {v + {m^2}} }}} \right.
 \kern-\nulldelimiterspace} {\sqrt {v + {m^2}} }})
\end{equation}
\begin{equation}
\sigma  = \sqrt {\log ({v \mathord{\left/
 {\vphantom {v {{m^2} + 1}}} \right.
 \kern-\nulldelimiterspace} {{m^2} + 1}})}
\end{equation}
Where $m$ and $v$ are the mean and variance of a lognormal random variable respectively, $\mu$ and $\sigma$ are the mean and standard deviation of the associated normal distribution respectively.

Step 3: Generate the traffic value of every base station in the region at each time using lognormal distribution with parameters $\mu (t)$ and $\sigma$ expressed as follows:
\begin{equation}
{V_i}(t) = lognrnd(\mu (t),\sigma )
\end{equation}
Where ${V_i}(t)$ is the traffic volume of the $i$th base station at time $t$.
Using this method, the traffic pattern of single base station in a typical region can be obtained and this model can be used for system analysis.
\subsection{Performance evaluation}
\begin{figure}[!t]
\centering
\includegraphics[width= 9cm]{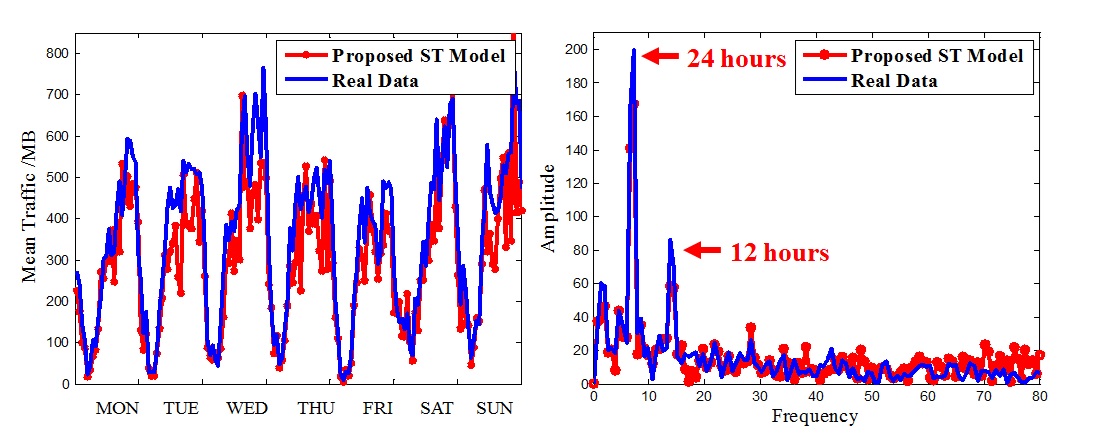}
\caption{(Left) Comparison between real data and proposed S-T model in time domain. (Right) Comparison between real data and proposed S-T model in frequency domain}
\label{fig_10}
\end{figure}

To evaluate the accuracy of the proposed spatial-temporal (S-T) traffic model, the real traffic data in park region and the traffic data generated by the model is compared as shown in Fig.~\ref{fig_10}. Fig.~\ref{fig_10} (Left) illustrates the mean traffic variation of real-world base stations in park region and of data obtained by the proposed S-T model in time domain. The results show that the mean square error between the proposed S-T model and real data is 5.5\%, which verified the accuracy of the model. Fig.~\ref{fig_10} (Right) demonstrates the main frequency components of real traffic data and the proposed S-T model are exactly the same. They both have periods of 24 hours and 12 hours. Therefore, the proposed S-T modeling approach can accurately match the traffic pattern in real cellular networks both in time domain and frequency domain.
\section{Conclusion}
\label{sec:conc}
The inhomogeneity of users and traffic is one of the important characteristics of heterogeneous cellular networks. The study of models which can accurately describe traffic patterns is necessary for designing and analyzing future ultra-dense networks. Firstly, a sinusoid superposition model is proposed to model the temporal traffic pattern of multiple base stations in a region. Secondly, the spatial distributions of traffic in several scenarios are analyzed including whether removing hotspots and in typical regions such as park, campus and CBD. It shows that lognormal distribution can be used to model spatial traffic distribution generally and the parameter $\sigma $  of the spatial traffic model is related to the type of typical regions. The empirical values of   $\sigma $ in three typical regions are given. Thirdly, a spatial-temporal traffic modeling approach for single base station is proposed. This model reflects both the randomness and periodicity of single base station's traffic variability. All the three models are evaluated and verified by comparing with the traffic data of s real cellular network in a big city of China.

In future, we will further study the universality of the proposed models and analyze the data of more typical regions in real cellular network. Additionally, the application of these models to estimate the energy consumption in future new network architectures will be investigated. The research of new energy saving algorithms is also possible based upon these traffic models.

\section*{Acknowledgment}

This work is supported by National 973 Program under grant 2012CB316005, the National Science Foundation of China (NSFC) under grant 61372114, the Fundamental Research Funds for the Central Universities under grant 2014ZD03-01, the New Star in Science and Technology of Beijing Municipal Science \& Technology Commission (Beijing Nova Program: Z151100000315077), the Beijing Higher Education Young Elite Teacher Project under grant YETP0434, the joint research between BUPT and CTBRI.

\end{document}